\documentclass[a4paper,11pt,5p]{elsarticle}
\usepackage{siunitx}
\usepackage{multirow}		%Needed for multi-row cells in tables
\usepackage{ctable}			%Needed for \toprule etc in table
\usepackage{threeparttable}
\usepackage{multicolumn, blindtext}
\usepackage{ulem} %for striking through text 
\usepackage{csquotes}		%For good quotations

\usepackage{lineno}
	\modulolinenumbers[5]
\biboptions{sort&compress}	
\begin{document}

\begin{frontmatter}

	\title{Adhesive Transfer operates during Galling}

		\author[IC]{Samuel R. Rogers\corref{cor1}}
		\ead{srr13@ic.ac.uk}
		\author[NU]{Jaimie Daure}
		\author[NU]{Philip Shipway}
		\author[RR]{David Stewart}
		\author[IC]{David Dye}
		\cortext[cor1]{Corresponding author}
		\address[IC]{Imperial College, South Kensington, London SW7 2AZ, UK}
		\address[NU]{University of Nottingham, University Park, Nottingham NG7 2RD, UK}
		\address[RR]{Rolls-Royce plc, Raynesway, Derby DE21 7WA, UK}

	\begin{abstract}
		In order to reduce cobalt within the primary circuit of pressurised water reactors (PWR's), wear-resistant steels are being researched and developed. In particular interest is the understanding of galling mechanisms, an adhesive wear mechanism which is particularly prevalent in PWR valves. Here we show that large shear stresses and adhesive transfer occur during galling by exploiting the 2 wt.\% manganese difference between 304L and 316L stainless steels, even at relatively low compressive stresses of \SI{50}{\mega\pascal}. Through these findings, the galling mechanisms of stainless steels can be better understood, which may help with the development of galling resistant stainless steels.
	\end{abstract}
	
\end{frontmatter}

%\section{Introduction}
	 There has been renewed interest in nuclear power generation in recent years, in an effort to reduce carbon emissions and reliance on fossil fuels. With new regulations and a desire to reduce cobalt in pressurised water reactor (PWR) primary circuits \cite{ONRCobalt} alternative materials are required. Of particular interest is the replacement of Stellite 6 (a cobalt hardfacing alloy) in valve seatings with a galling-resistant stainless steel alloy. %Despite considerable efforts in developing an iron-based hardfacing alloy, little work has been produced on the fundamental mechanisms of galling within stainless steels.

	Galling is an adhesive wear mechanism which is known to result in severe surface degradation, and may result in sliding surfaces seizing \cite{UnOxVsOx,ASTMG98,ASTMG196}. Whilst self-mated stainless steel is well documented to show poor galling resistance, particularly at elevated temperature, considerable efforts have been undertaken to develop a galling resistant stainless steel or iron-based hardfacing alloy \cite{UnOxVsOx,StainlessSteels,304Galling,SchumacherStrainHardening,AustralianStainlessSteel,GallingCanada,FrenchGalling,RTOcken,KimKim,GallTough,SmithThesis,NitromaxxMasters,BowdenThesis,GallingTorque}.	
	
	Work has been produced which elucidates the galling mechanisms which may occur, many of these have been on self-mated test pairs, making adhesive transfer difficult to discern. Whilst a number of non self-mated tests have also been performed, the two materials are often found to differ quite considerably, be it through differing hardness, yield strengths, phases, microstructures or widely differing chemistries \cite{StainlessSteels,ReidThesis,GallingCanada,SteelersGalling}. This work seeks to address these issues, by performing non self-mated tests using two very mechanically similar stainless steels but which differ sufficiently in chemistry (namely molybdenum content) in order to observe any adhesive transfer which may have occurred.

	\begin{figure}[h]
	    	\centering
	    	\includegraphics[width=8cm]{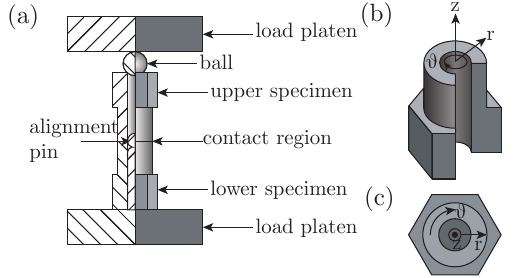}
	    	\caption{(a) ASTM G196 galling rig, redrawn from \cite{ASTMG196}; (b) ASTM G196 galling sample, with a section removed to enable a view of the radial cross-section; (c) top view of an ASTM G196 galling sample.}
	    	\label{G196}
	    \end{figure}
	    
	    Galling tests were performed using the ASTM G196 method,  Figure \ref{G196}, under atmospheric conditions at the University of Nottingham. An applied compressive load of  \SI{50}{\mega\pascal} was used. \SI{50}{\mega\pascal} was chosen since this would ensure that galling took place between the surfaces. Tests used non self-mated pairs of 314L/316L stainless steel. Virgin surfaces were used for each test. Surfaces had been ground to a finish of $\pm$ \SI{10}{\micro\meter}.
	    
	    Upon the completion of testing, the samples' surfaces were scanned using white light interferometry (WLI) before being processed using specially written Python code to process the surfaces (\text it{tilt correct, remove edge effects and interpolate}). Once processed, surfaces were quantified, using the maximum \& minimum surface heights, R$_{t}$ (max. + min.), galled area and volume change. These quantities are described in \cite{UnOxVsOx,GallingBook}.

	Samples were sectioned and prepared for metallographic examination using SEM in both secondary electron (SE) and backscattered electron (BSE) imaging modes. Energy dispersive X-ray spectroscopy (EDX) was also used to form elemental maps of cross-sections.

%\section{Results \& Discussion}

    \begin{figure*}[h!]
			\centering
			\includegraphics[width=15.5cm]{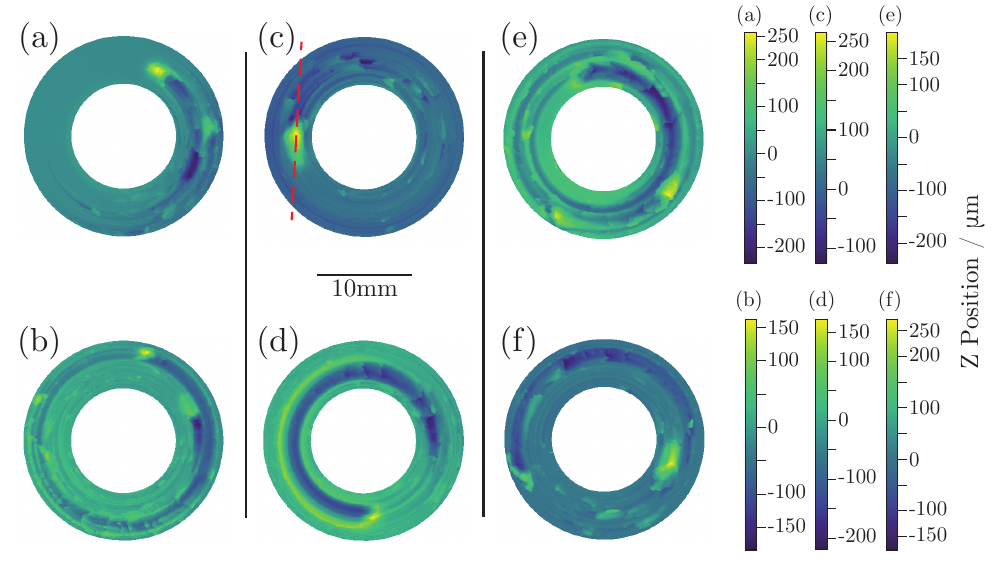}
			\begin{threeparttable}
			\centering
			%\caption{Quantification of the galling damage using surface profile measurements for samples shown in Figure \ref{WLI} .}
			\label{MeasuringGalling}
			%\begin{tabular}{m{3.5em}m{6em}m{2.5em}m{2.5em}m{2.5em}} %l=left justified; m=middlealigned; >{\raggedright} for not justifying text - USE THIS ONE TO HAVE MORE CONTROL OVER COLUMN SIZES (at least, it works...)
			{\small
				\begin{tabular}{ccccccc}
				\toprule
				\toprule
					\multirow{2}{3.5em}{\textbf{Sample}}		&	\multirow{2}{8em}{\textbf{Applied Stress / \SI{}{\mega\pascal}}}        & \multicolumn{3}{c}{\textbf{Sample Height / \SI{}{\micro\meter}}}   &   \multirow{2}{5em}{\textbf{Galled Area / \%}} & \multirow{2}{5em}{\textbf{$\Delta$V / \SI{}{\milli\meter\cubed}}} \\
				&	&	   \textbf{Max}   & \textbf{Min}   & \textbf{Rt}  & &   \\
					
				\midrule
					(a)	& 50   & 257 &   -233 &   490 &   12.04   &   -0.2440 \\
					(b)	& 50 &   162 &   -182 &   344 &   18.61   &   -0.0916 \\
					(c)	& 75 &   262 &   -126 &   388 &   19.75   &   +0.3068 \\
					(d)	& 75 &   171 &   -219 &   390 &   18.87   &   -0.3005 \\
					(e)	& 100 &   198 &   -237 &   435 &   21.69   &   -0.2533 \\
					(f)	& 100 &   270 &   -175 &   445 &   16.82   &   +0.0121 \\
				\bottomrule
				\end{tabular}
				}
		\end{threeparttable}

			\caption{White light interferometry height maps of 316L (a), (c) \& (e) vs 304L (b), (d) \& (f) stainless steel samples tested at various loads in the non-oxidised conditions, with their corresponding height scales and quantification measures. Tests were paired in the following way: (a) \& (b), (c) \& (d) and (e) \& (f). The red dashed line on sample (c) denotes the cross-section which was used to image Figures 3 and 4.}
			\label{304V316Surfaces}
		\end{figure*}

	When the galling damage of non self-mated tests, Figure \ref{304V316Surfaces}, are compared with that of self-mated 316L galling tests at the same applied stress, it can be seen that the extent of the galling damage is very similar \cite{UnOxVsOx}, confirming the validity of using 304L as a mating pair with 316L stainless steel. All samples tested show significant galling as a result of their high applied stresses. Whilst 3 different stresses were tested, the quantitative measures of galling for each test pair are similar, meaning that the effect of loading on the extent of galling is inconclusive.
    
    %It can be noted that across the 3 tests, the sample heights (maximum, minimum and R$_{t}$) and galled area are seen to be very similar across all samples, Figure \ref{304V316Surfaces}.
    
    Since the damage morphology is consistent with that seen in self-mated 316L galling tests \cite{UnOxVsOx}, it can be concluded that the wedge formation and flow mechanism (Archard adhesive wear mechanism) has also occurred for 304L/316L non self-mated tests, Figure \ref{304V316Surfaces}. A full description for the wedge formation and flow mechanism (also known as the Archard adhesive wear theory) can be found in \cite{UnOxVsOx}. In each non self-mated test, the damage morphology appears consistent with a single surface providing most of the material for the galling peak. In the case of Figure \ref{304V316Surfaces} surface (a) and (b), two instances of this have occurred, with each surface appearing responsible for the growth of a galling peak. In addition to material transfer, material has also been lost from the system in this test, as given by the negative $\Delta$V for both surfaces. The test between surface (c) and (d) of Figure \ref{304V316Surfaces} shows a single instance of galling to have occurred, with surface (d) primarily giving material for adhesion junction growth. It can be seen that the all of the material displaced from surface (d) was adhesively transferred to surface (c), as shown by their almost identical changes in volume, Figure \ref{304V316Surfaces}. The final test, between surfaces (e) and (f) show a primary galling scar, however, there are a number of smaller scars which are also seen. Although adhesive transfer is seen to occur from surface (e) to (f), material has been lost from the system.
 
	\begin{figure}[h]
	    	\centering
	    	\includegraphics[width=9cm]{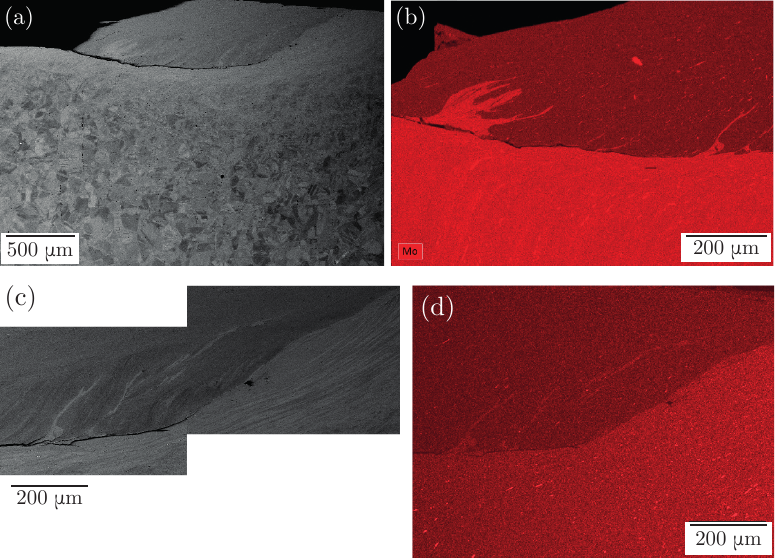}
	    	\caption{A circumferential cross-section of a galling peak of 316L stainless steel, Figure \ref{304V316Surfaces}(c), from a non self-mated test of 304 vs 316L stainless steel sample tested at a normal load of \SI{75}{\mega\pascal}. (a) The whole peak as viewed using SE imaging; (b) Mo EDX map of the peak front; (c) peak rear as viewed using SE imaging; (d) Mo EDX map of the peak rear.}
	    	\label{NSMMain}
	    \end{figure}
    
   In order to verify if adhesive transfer occurred between the non self-mated samples, they needed to be sectioned in order to view the samples sub-surface. Figure \ref{NSMMain} shows a portion of a circumferential cross-section of the surface shown in Figure \ref{304V316Surfaces}(c), which is a 316L stainless steel sample, and shows the main galling peak on the sample. As was seen for self-mated 316L mated pairs, the underlying microstructure is clearly distinguished from the peak and the tribologically affected zone (TAZ) \cite{UnOxVsOx}. %What is less clear at this magnification is the deformed region below the TAZ, however, this is made clear at higher magnifications, \textit{e.g.} in Figure ???.
    
    It is interesting to note that the the galling peak is adhered to the underlying material at the peak rear, Figure \ref{NSMMain}(a) \& (c), whilst being seemingly unattached at the peak front. This suggests that the adhesion junction formed at the peak rear, and although adhesion may have also occurred at the peak front, the adhesion bond in this region was not as strong as at the original adhesion junction.
    
    By performing an EDX scan over the front portion of the galling peak, it can be seen that the predominant material of the peak is 304L stainless steel, as evidenced by its low Mo concentration, Figure \ref{NSMMain}(b), meaning that adhesive transfer of a significant volume has occurred. In addition, a portion of 316L appears to be mechanically mixed within the 304L peak, Figure \ref{NSMMain}(b). Whilst from this section alone, it is difficult to know how this material got there, by observing the mechanical mixing on the right of Figure \ref{NSMMain}(b), it can be seen that a small amount of the 316L substrate appears mixed into the 304L peak. It may therefore be that the larger portion of 316L within the peak has been mechanically mixed into the 304L. It is difficult to give certainty on this, however, since it could be that the 316L first transferred over to the 304L mating surface, before being mechanically mixed and transferred back to the the 316L mating surface.
    
    When observing the rear of the galling peak, Figure \ref{NSMMain}(c) \& (d), further portions of mechanically mixed 316L within the 304L peak can be seen. These are finer in width and longer, suggesting that they have undergone a greater extent of shear, and potentially mechanical mixing, than those towards the front of the galling peak, Figure \ref{NSMMain}(b). Flow lines are much more visible within both the galling peak and the substrate material at the peak rear.
        
        \begin{figure}[t!]
	    	\centering
	    	\includegraphics[width=9cm]{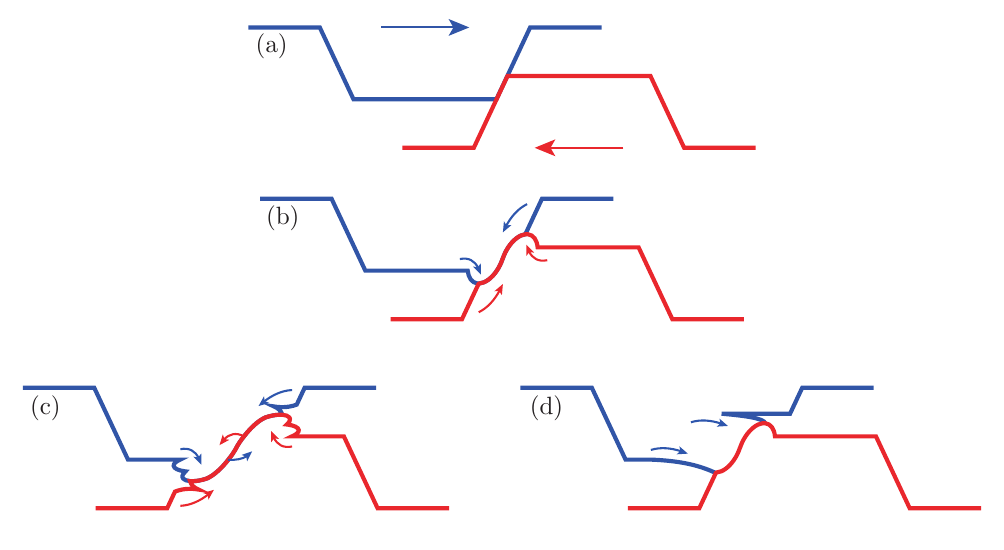}
	    	\caption{A refinement of the Archard adhesive wear mechanism (wedge formation and growth mechanism) \cite{UnOxVsOx}, with the addition of material flow, as described using colour co-ordinated arrows based on the observation from this work.}
	    	\label{RefinedMech}
    	\end{figure}

    With this information, we can refine the Archard adhesive wear mechanism \cite{UnOxVsOx}, in order to include mechanical mixing within and between tribosurfaces, Figure \ref{RefinedMech}. Two asperities come into contact and form an adhesive junction, Figure \ref{RefinedMech}(a), subsequent shearing of this junction results in wedge formation (this can also occur through the shearing of two flat surfaces that have adhered), Figure \ref{RefinedMech}(b). At this point, the mechanism can continue in one of two ways: Figure \ref{RefinedMech}(c), material from both tribosurfaces causes wedge growth to such an extent that excess material ahead of the prows folds over, whilst shear failure occurs behind the prow, resulting in the formation of lips, or; Figure \ref{RefinedMech}(d) material from one tribosurface preferentially supplies material for wedge growth, resulting in a very large wedge on one tribosurface and a considerably smaller wedge on the opposing surface. The mechanism as shown in Figure \ref{RefinedMech}(d) was found to primarily occur in this work. Whilst appearing consistent for the stainless steel pairings tested in this work and in \cite{UnOxVsOx}, this mechanism may be one of a number of galling mechanisms. In addition, further work is needed to fully understand galling initiation.

        \begin{figure}[h]
	    	\centering
	    	\includegraphics[width=9cm]{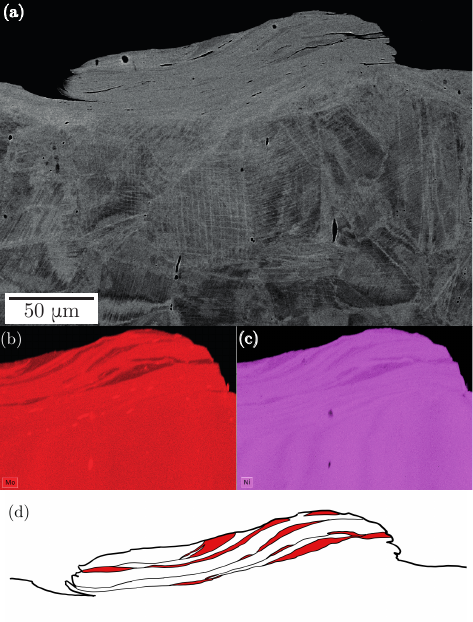}
	    	\caption{(a) A backscattered electron image and EDX maps of (b) Mo and (c) Ni of a small galling peak on the same circumferential section as viewed in in Figure \ref{NSMMain}. The scale bar is common across (a) - (c). (d) A sketch of the galling peak shown in (a), showing the layers of 316L and 304L stainless steel that have adhesively transferred through successive galling events to form the final galling peak which is observed. Red portions of the sketch are 304L stainless steel, with uncoloured regions being 316L stainless steel.}
	    	\label{NSMSmallPeak}
    	\end{figure}
    
    Across the circumferential cross-section viewed in Figure \ref{NSMMain}, an additional smaller galling peak can be also be seen, Figure \ref{NSMSmallPeak}. If this small peak is mapped using EDX, it can be seen that unlike in Figures \ref{NSMMain}, the entirety of the peak appears to be made up of a number of mechanically mixed regions of 304L and 316L stainless steels, Figure \ref{NSMSmallPeak}. Given that the regions of 304L seem to be isolated within the 316L material within the peak, that all regions of 304L appear relatively thin and long, and that the volume of 316L material within the peak does not appear to have all come from this galling trough, it appears an alternative explanation for the galling peak morphology is required. The most likely possibility is that multiple instances of galling and adhesive transfer have occurred, resulting in a galling peak with a layered internal structure, Figure \ref{NSMSmallPeak}(d), and when considering the volume of 316L, a peak which is considerably larger than the associated trough size would suggest.

%\section{Conclusions}
	In summary, as with self-mated 316L stainless steel, non self-mated 304L vs 316L galling pairs appear to gall via the wedge formation and growth, or Archard adhesive wear, mechanism. Due to chemical variations between the two alloys (namely the incorporation of molybdenum in 304L), adhesive transfer was observed between surfaces. Mechanical mixing between surfaces was also observed, in addition to the previously observed mechanical mixing in a single surface. A series of smaller galling peaks were observed to have formed, likely through successive galling and adhesive transfer events.

\section*{Acknowledgements}
We gratefully acknowledge support from Rolls-Royce plc, from EPSRC (EP/N509486/1) and from the Royal Society (D. Dye Industry Fellowship). The authors gratefully acknowledge funding from Rolls-Royce plc to the University of Nottingham in support of this work.

\section*{References}	%* denotes lack of number

\end{document}